\documentclass{article}

\usepackage{epsfig}
\input epsf.tex
\usepackage[english]{babel}
\usepackage{color}
\newcommand{\parl}{\parallel}
\usepackage{graphicx}
\usepackage{amsmath}
\usepackage{subcaption}

\begin{document}

\title{Quantum Confined Stark Effect in Wide Parabolic Quantum Wells}
\author{Sylwia Zieli\'{n}ska-Raczy\'{n}ska, Gerard Czajkowski, and David
Ziemkiewicz \\
\\
\emph{UTP University of Science and Technology,
Bydgoszcz},\\\emph{ Al. Prof. S. Kaliskiego 7, 85-789 Bydgoszcz,
\emph{Poland}}} \maketitle
\begin {abstract}

We show how to compute the optical functions of Wide Parabolic
Quantum Wells (WPQWs) exposed to uniform electric \textbf{F}
applied in the growth direction, in the excitonic energy region.
The effect of the coherence between the electron-hole pair and the
electromagnetic field of the propagating wave including the
electron-hole screened Coulomb potential is adopted, and the
valence band structure is taken into account in the cylindrical
approximation. The role of the interaction potential and of the
applied electric field, which mix the energy states according to
different quantum numbers and create symmetry forbidden
transitions, is stressed. We use the Real Density Matrix Approach
(RDMA) and an effective e-h potential, which enable to derive
analytical expressions for the WPQWs electrooptical functions.
Choosing the susceptibility, we performed numerical calculations
appropriate to a  GaAs/GaAlAs  WPQWs.  We have obtained a red
shift of the absorption maxima (Quantum Confined Stark Effect),
asymmetric upon the change of the direction of the applied field
(\textbf{F}$\to$ -\textbf{F}), parabolic for the ground state and
strongly dependent on the confinement parameters (the QWs sizes),
changes in the oscillator strengths, and new peaks related to the
states with different parity for electron and hole.
\end {abstract}
\section{Introduction}
The effects on optical spectra when an external electric field is
applied, known in atomic physics as the Stark Effect, evolved very
rapidly with the invention and development of semiconductor
nanostructures. The effects of confinement of carriers overlap
with the interaction with the field giving rise to the new
phenomenon known as the Quantum Confined Stark Effect (QCSE).
First reported for Quantum Wells by Miller \emph{et all}.
\cite{MillerChemla1984}, \cite{MillerChemla1985}, the QCSE is
continuously attracting the interest of the researches. The
references \cite{MillerChemla1984} -\cite{Wilkes} are only a small
collection of a very large number of papers studying the
properties of various nanostructures (Quantum Wells, Quantum Dots,
Quantum Rods, Superlattices etc.) under electric field. In most of
this nanostructures the applied electric field causes a red-shift
of the positions of the lowest energy states, changes in the
exciton binding energy, and lowering the oscillator strengths of
the resonances.  Here we consider QCSE in Wide Parabolic Quantum
Wells (WPQWs), of thicknesses in the growth direction of the order
of a few excitonic Bohr radii of the well material (see, for
example, \cite{Tabata2007}-\cite{EPJ.2015} and references
therein). The optical spectra of WPQWs show a large number of
resonances, which are due to the transitions between confined
states. The Coulomb e-h potential and different confinements for
electrons and holes cause mixing of the states with different
quantum (confinement) numbers. When additionally an electric field
is applied, states symmetry forbidden appear in the spectra. The
behavior of the positions of the resonances is more complicated
than in the narrow QWs since the lower states show a red-shift,
but some higher states show a blue-shift or a zig-zag shape, and
their oscillator strengths decrease. The case without the electric
field was discussed in our previous paper~\cite{EPJ.2015}. Using the same
model, we are able to include the effects of the applied field and
obtain the solution in analytical form. As an example, we consider
a WPQW with GaAs as the optically active layer and
Ga$_{1-x}$Al$_x$As as the barriers, where the active layer is of
the extension of a few excitonic Bohr radii, and the constant
electric field is applied in the growth direction, which we
identify with the $z$ axis.

Although our investigations deal with the theoretical model of
WPQW exposed to uniform electric field it is believed that such
systems are important due to their controllability and potential
applications. Employing an external electrostatic field to quantum
well allows one for steering the optical properties of the system.
Together with the geometric characteristic of QW the external
field is one of the strong modulating factor influencing the
energy spectrum of charge carriers. Due to controllability of the
field the optical properties of the nanostructures can be changed
on demand. Performing the manipulations of the external
interaction on WPQW
 gives one  possibility
of an effective processing of electrosusceptibility,
which may in the future be exploited for constructing electrooptical modulators or
optoelectronic processors.

Our paper is organized as follows. In the section 2,
 we present the assumptions of considered model and solve the
 constitutive equation with effective electron-hole interaction potential and the applied field.
 Section 3 is devoted to the details of the applied potential. Next, in section 4,
  the derived solution of constitutive equation is used to obtain the energy levels of the considered
  GaAs/Ga$_{1-x}$Al$_x$As
  wide parabolic quantum well. Finally, in section 5, the electrosusceptibility for such nanostructure is calculated and discussed.

\section{The Model}
We will compute the linear optical response of a WPQW to  a plain
electromagnetic wave
\begin{equation}\label{stpar_wave}
E_i(z,t)=E_{i0}\exp({\rm i}k_0z-{\rm i}\omega t), \qquad
k_0=\frac{\omega}{c},
\end{equation}
\noindent attaining the boundary surface of the WPQQW active layer
located at the plane  $z=-L/2$. The second boundary is located at
the plane
 $z=L/2$. The movement of the carriers in the $z$ direction is determined by one-dimensional parabolic potentials,
 characterized by the oscillator energies $\hbar\omega_e,\hbar\omega_h$, respectively. Additionally, an external electric field
  \textbf{F} is applied parallel to the
 $z$ axis.  We adopt the real density matrix
approach (RDMA) to compute the optical properties (see, for
example, \cite{StB87}-\cite{RivistaGC}). In this approach the
linear optical response will be described by a set of coupled
equations: two constitutive equations for the coherent amplitudes
$Y_\nu(\textbf{r}_e,\textbf{r}_h)$, $\nu=\hbox{H,L}$ stands for
heavy-hole (H) and light-hole exciton); from them the polarization
can be obtained and used in Maxwell's field equations.  Having the
field we can determine the QW electroptical functions
(reflectivity, transmission, and absorption).

Thus the next steps are the following: We formulate the
constitutive equations. The equations will be then solved giving
the coherent amplitudes $Y$. From the amplitudes we compute the
polarization inside the Quantum Well, the electric field of the
wave propagating in the QW, and the optical functions.


 The constitutive equation
for the coherent amplitude
 $Y$ in a WPQW and with the applied homogeneous electric field $\textbf{F}=F\textbf{k}$ has the
 form (see, for example, \cite{RivistaGC})
\begin{eqnarray}\label{stpar_konstytutpow}
&&\biggl[E_{g}-\hbar\omega-{\rm
i}{\mit\Gamma}+\frac{\hat{p}_{ez}^2}{2m_{e}}+\frac{\hat{p}_{hz}^2}{2m_{hz}}+\frac{\hat{\bf
p}_{\rho}^2}{2\mu_{\parl }}+\frac{\hat{\bf p}_{\parl}^2}{2M_{\parl
}}+eFz_e-eFz_h+V_{eh}(\rho,z_e,z_h)\nonumber\\
&&+
\frac{1}{2}m_e\omega_e^2z_e^2+\frac{1}{2}m_{zh}\omega_h^2z_h^2\biggr]Y
= {\bf M}({\bf r}){\bf E}({\bf R}),
\end{eqnarray}
\noindent where $\rho=\sqrt{(x_e-x_h)^2+(y_e-y_h)^2}$ is the
two-dimensional e-h distance,  $V_{eh}(\rho,z_e,z_h)$ is the
electron-hole interaction potential, ${\bf M}({\bf r})$ is the
transition dipole density, which form we have assumed as
\begin{eqnarray}\label{dipoledensity}
&&{\bf M}({\bf r})={\bf M}(\rho,z,\phi)
=\frac{\textbf{M}_{0}}{2\pi\rho_{0}}\delta(z)\delta\left(\rho-\rho_{0}\right),
\end{eqnarray}
\noindent  $z=z_e-z_h$ being the relative coordinate in the $z$
direction, $\rho_{0}$ is the coherence radius (the physical
meaning was explained, for example, in \cite{StB87},
\cite{CBT96}), ${\bf R}$ jest is the excitonic center-of-mass
coordinate, ${\bf E}({\bf R})$ is the electric field vector of
the wave propagating in the QW;
 and  $\hat{\bf p}_{\rho}$, $\hat{\bf p}_{\parl}$  are
the momentum operators for the excitonic relative- and
center-of-mass motion in the QW plane. In the consideration of
narrow QWs (with extension less than one excitonic Bohr radius and
arbitrary confinement shape) the following approximation was often
used. The movement in the $z$- direction was decoupled from the
movement in the $xy$ plane, and the electron-hole interaction was
assumed in the 2-dimensional form
\begin{equation}
V_{eh}=-\frac{1}{4\pi\epsilon_0\epsilon_b}\frac{e^2}{\rho}
\end{equation}
with the QW material dielectric constant $\epsilon_b$. This
approximation enabled to obtain analytical solutions for the
electron and the hole wave functions, and thus the calculation of
the optical properties (see for example,\cite{RivistaGC}). Such
metod cannot be used in the considered case of wide QWs (the
extension of several excitonic Bohr radii) since the e-h
interaction retains its 3-dimensional character. As was pointed in
Refs. \cite{arxiv}, \cite{EPJ.2015}, the direct numerical solution
of the constitutive equation (\ref{stpar_konstytutpow}) is, at the
moment, not available because of lack of the appropriate
orthonormal basis to use in order to decrease the dimension of the
6-dimensional configuration space, \cite{Schillak_epj}. Therefore
we use the following 3-dimensional form of the interaction
potential
\begin{equation}\label{stpar_potencjalharmoniczny}
V_{eh}=-S\exp\left[-v\left(z_e-z_h\right)^2-w\rho^2\right].
\end{equation}
\noindent with parameters  $v,w$  appropriate  for a given
nanostructure, which enables to perform analytical calculations
and reproduces the basic properties of the exciton
(\cite{arxiv},\cite{EPJ.2015}).

 In the following we
assume that the propagating wave is linearly polarized in the $x$
direction, and that the vector \textbf{M} has a non-vanishing
component in the same direction.  We find in the equation
(\ref{stpar_konstytutpow}) Hamilton operators for the
one-dimensional harmonic oscillator
\begin{equation}
\hat{H}_e=\frac{\hat{p}_{ez}^2}{2m_e}+\frac{1}{2}m_e\omega_e^2z_e^2+eFz_e,
\end{equation}
\noindent with an analogous expression for $\hat{H}_h$. Therefore
we look for a solutions $Y$ in terms of the eigenfunctions and eigenenergies of the
operators $H_e,H_h$:
\begin{eqnarray}\label{stpar_eigenf1doscillator8}
\psi_{ej}(\xi_e)&=&N_{ej}e^{-\xi_e^2/2}H_{j}(\xi_e)=\vert {\rm e}j\rangle,\nonumber\\
\xi_e&=&\alpha_e z_e-a_e,\qquad\alpha_{e} = \sqrt{\frac{m_{ez} \omega_{e}}{\hbar}}\nonumber\\
a_e&=&-\frac{1}{2\alpha_e^3}\left(\frac{2m_e}{\hbar^2}eF\right)\\
\label{starkshift} E_{je}&=&\frac{\hbar\omega_e}{2}(2j+1)
-\frac{\hbar\omega_e}{8\alpha_e^6}\left(\frac{2m_e}{\hbar^2}eF\right)^2
\end{eqnarray}
with analogous expressions for the hole, where
\begin{eqnarray}
\xi_h&=&\alpha_hz_h+a_h,\quad \alpha_{h} = \sqrt{\frac{m_{hz} \omega_{h}}{\hbar}},\nonumber\\
a_h&=&-\frac{1}{2\alpha_h^3}\left(\frac{2m_h}{\hbar^2}eF\right),
\end{eqnarray}
  $H_j(x)$ are Hermite polynomials and $N_j$ normalization
  constants. Taking into account the valence band structure and
  heavy- and light holes excitons, we obtain
\begin{eqnarray}
a_h&\to&a_{hH},\nonumber\\
a_{hH}&=&-\frac{1}{2\alpha_{hH}^3}\left(\frac{2m_{hzH}}{\hbar^2}eF\right)
=-\frac{1}{2\alpha_{hH}^3a^{*3}_{H}}\left(\frac{2m_{hzH}}{\hbar^2}ea^{*3}_{H}F\right)\\
&=&-\frac{1}{2\tilde{\alpha_{hH}}^3}\left[\left(\frac{m_{hzH}}{\mu_{\parallel
H}}\right) \left(\frac{2\mu_{\parallel
H}}{\hbar^2}ea^{*3}_{H}F\right)\right]
=-\frac{1}{2\tilde{\alpha_{hH}}^3}\left(\frac{m_{hzH}}{\mu_{\parallel
H}}\right)\frac{F}{F_{{\rm I}H}},\nonumber
\end{eqnarray}
where
\begin{equation}
\tilde{\alpha}_H=\alpha_H a^*_H,
\end{equation}
and $F_{{\rm I}H}$ is the so-called ionization field
\begin{equation}
F_{{\rm I}H}=\frac{\hbar^2}{2\mu_{\parallel
H}}ea^{*3}_H=\frac{R^*_H}{ea^*_H}.
\end{equation}
For the GaAs data we have $F_{{\rm I}H}=2.318~{\rm kV/cm}$ and
$F_{{\rm I}L}=3.286~{\rm kV/cm}$. Analogously, we have
\begin{equation}
a_e=-\frac{1}{2\tilde{\alpha}_e^3}\left(\frac{m_e}{\mu_{\parallel
H}}\right)\frac{F}{F_{{\rm I}H}}.
\end{equation}
The expression for the Stark shift in the equations
(\ref{starkshift}) takes now the form
\begin{eqnarray}
&&-\frac{\hbar\omega_e}{8\alpha_e^6}\left(\frac{2m_e}{\hbar^2}eF\right)^2=-\frac{\hbar\omega_e}{8\tilde{\alpha}_e^6}\left(\frac{m_e}{\mu_{\parallel
H}}\right)\left(\frac{F}{F_{{\rm I}H}}\right)^2,\nonumber\\
\end{eqnarray}
and
\begin{eqnarray}
&&-\frac{\hbar\omega_{hH}}{8\alpha_{hH}^6}\left(\frac{2m_{hzH}}{\hbar^2}eF\right)^2=
-\frac{\hbar\omega_{hH}}{8\tilde{\alpha}_{hH}^6}\left(\frac{m_{hzH}}{\mu_{\parallel
H}}\right)\left(\frac{F}{F_{{\rm I}H}}\right)^2.
\end{eqnarray}
For the light-hole excitons we obtain quite analogous expressions,
substituting the respective parameters. Using the above functions,
we seek the solution for $Y$ in the form
\begin{equation}\label{stpar_rozwiniecie1}
Y\left(\rho,\xi_e,\xi_h\right)=\sum\limits_{j,n=0}^{N}\psi_{ej}(\xi_e)\psi_{nh}(\xi_h)Y_{jn}(\hbox{\boldmath
$\rho$})=\sum\limits_{j,n=0}^{N}Y_{jn}(\hbox{\boldmath
$\rho$})\vert{\rm e}j{\rm h}n\rangle,
\end{equation}
Substituting (\ref{stpar_rozwiniecie1}) into the eq.
 (\ref{stpar_konstytutpow}) we obtain equations for the functions
 $Y_{jn}$
\begin{eqnarray}\label{stpar_konstytutpow1}
\sum\limits_{j,n=0}^{N}\left[E_{g}-\hbar\omega-{\rm
i}{\mit\Gamma}+E_{je}+E_{nh}+\frac{\hat{\bf p}_{\hbox{\boldmath
$\rho$}}^2}{2\mu_{\parl }}+\frac{\hat{\bf p}_{\parl}^2}{2M_{\parl
}}+V_{eh}(\hbox{\boldmath $\rho$},\xi_e,\xi_h)
\right]\psi_j(\xi_e)\psi_{n}(\xi_h)Y_{jn}(\hbox{\boldmath $\rho$})
= {\bf M}({\bf r}){\bf E}({\bf R}).\nonumber\\
\end{eqnarray}
 We assume the so-called long-wave
approximation and consider ${\bf E}({\bf R})$ in the equation
 (\ref{stpar_konstytutpow1}) as a constant quantity.

 \begin{table}[ht]
\caption{\small Band parameter values for GaAs, AlAs, and
Ga$_{0,7}$Al$_{0,3}$As,
 AlAs data from~\protect\cite{GrundmannStier}, for Ga$_{0.7}$Al$_{0.3}$As
by linear interpolation. Energies in meV, masses in free electron
mass $m_0$, $\gamma_1, \gamma_2$ are Luttinger parameters}
\label{parametervalues}
\begin{small}
\begin{center}
\begin{tabular}{c c c c c}
\hline\\
Parameter &GaAs  &  & AlAs&Ga$_{0.7}$Al$_{0.3}$As \\
\hline
$E_g$ & 1519.2&  & 3130 &2002\\
$m_e$     &0.0665 &  &0.124 &0.084\\
$\gamma_1$&6.85&&3.218&\\
$\gamma_2$&2.1&&0.628&\\
$m_{h\parallel H}$ &0.112 &  &0.26&\\
$m_{h\parallel L}$ & 0.210  &  & 0.386& \\
$\mu_{\parallel H}$ &  0.042 &  & & \\
$\mu_{\parallel L}$ &  0.05 &  & & \\
$m_{hzH}$ & 0.38& & 0.51&0.39 \\
$m_{hzL}$ & 0.09 & & 0.22&0.13\\
$R^*_H$&3.64&&13.32&\\
$R^*_L$&4.3&&19.35&\\
$R^*_e$&5.76&&&\\
$a^*_H$&15.78&&7.03&\\
$a^*_L$&13.265&&4.84&\\
$a^*_e$&9.97&&&\\
$\epsilon_b$ & 12.53 &  & 11.16&12.12\\
\hline\\
\end{tabular}

\end{center}
\end{small}
\end{table}
  Using  the dipole density (\ref{dipoledensity}), the
model potential (\ref{stpar_potencjalharmoniczny}), and neglecting
the center-of-mass in plane motion, we put the constitutive
equation (\ref{stpar_konstytutpow}) into the form
\begin{equation}\label{stpar_konstytuwn3}
\left(E_{rs}+\frac{\hat{\bf p}_{\hbox{\boldmath
$\rho$}}^2}{2\mu_{\parl
}}\right)Y_{rs}-e^{-w\rho^2}\sum\limits_{nj}V_{rsnj}Y_{nj}=E\frac{M_0}{2\pi\rho_0}\langle
r\vert\delta(z_e-z_h) s\rangle\delta\left(\rho-\rho_0\right),
\end{equation}
where
\begin{eqnarray}\label{stpar_parzystosc}
E_{rs}&=&E_g+E_{re}+E_{sh}-\hbar\omega-{\rm
i}{\mit\Gamma},\qquad r,s,=0,1,2,\ldots\,,\nonumber\\
\label{stpar_elementymacierzowe} V_{rsnj}&=&S\langle {\rm e}r{\rm
h}s\left|\exp\left[-v\left(z_e-z_h\right)^2\right]\right\vert {\rm
e}n{\rm h}j\rangle.
\end{eqnarray}
\noindent When the electric field is absent, only states of the
same parity will give non vanishing elements $\langle r\vert
s\rangle$. For the field $F\neq 0$, due to the displacement
between the electron and hole confinement eigenfunctions, all
possible combinations, for example  $\vert 0e0h\rangle, \vert
0e2h\rangle, \vert 1e3h\rangle$, but also $\vert 1e0h\rangle$ etc.
have to be taken into account.
\section{Calculation of the constitutive equation coefficients}
We have to compute the elements $\langle r\vert\delta(z_e-z_h)
s\rangle$ and the potential matrix elements
\begin{eqnarray}
\langle {\rm e}r\vert\delta(z_e-z_h) {\rm h}s\rangle&=&\int\int
{\rm d}\xi_e{\rm
d}\xi_h\psi_{er}(\xi_e)\psi_{hs}(\xi_h)\delta(z_e-z_h)\nonumber\\
&=&\alpha_e\alpha_h\int\int {\rm d}z_e{\rm d}z_h\psi_{er}(\alpha_e
z_e-a_e)\psi_{hs}(\alpha_h z_h-a_h)\delta(z_e-z_h)\\
&=&\alpha_e\alpha_h\int {\rm d}z_h\psi_{er}(\alpha_e
z_h-a_e)\psi_{hs}(\alpha_h z_h-a_h).\nonumber
\end{eqnarray}
To perform this calculations, we use the transformation
\begin{eqnarray}\label{stpar_transformacja}
z_e-z_h=X,&& z_e+\frac{\alpha_h^2}{\alpha_e^2}z_h=Y,\nonumber\\
{\rm d}z_e{\rm
d}z_h&=&\frac{\alpha_e^2}{\alpha_e^2+\alpha_h^2}{\rm d}X{\rm d}Y
\end{eqnarray}
\noindent from which we have
\begin{eqnarray}\label{stpar_transformacjaodwotna}
z_e&=&\frac{\alpha_e^2}{\alpha_h^2+\alpha_e^2}\left(\frac{\alpha_h^2}{\alpha_e^2}X+Y\right),\nonumber\\
z_h&=&\frac{\alpha_e^2}{\alpha_h^2+\alpha_e^2}(Y-X).
\end{eqnarray}
\noindent Using the new variables we transform the expressions
\begin{eqnarray}
&&\xi_e^2+\xi_h^2+vX^2=\left(\alpha_ez_e-a_e\right)^2+\left(\alpha_hz_h-a_h\right)^2+vX^2\nonumber\\
&&=\left[\frac{\alpha_e^3}{\alpha_e^2+\alpha_h^2}\left(\frac{\alpha_h^2}{\alpha_e^2}X+Y\right)-a_e\right]^2
+\left[\frac{\alpha_h\alpha_e^2}{\alpha_e^2+\alpha_h^2}(Y-X)-a_h\right]^2+vX^2\nonumber
\end{eqnarray}
\noindent into
\begin{eqnarray}
&&\left(\frac{\alpha_e^2\alpha_h^2}{\alpha_e^2+\alpha_h^2}+v\right)X^2
-2\frac{\alpha_e\alpha_h}{\alpha_e^2+\alpha_h^2}\left(a_e\alpha_h+a_h\alpha_e\right)X\nonumber\\
&&+\frac{\alpha_e^4}{\alpha_e^2+\alpha_h^2}Y^2
-2\frac{\alpha_e^2}{\alpha_e^2+\alpha_h^2}\left(a_e\alpha_e-
a_h\alpha_h\right)Y+a_e^2+a_h^2.
\end{eqnarray}
\noindent Now the matrix elements (\ref{stpar_elementymacierzowe})
can be expressed by integrals
\begin{equation}
\int\limits_{-\infty}^\infty
Y^{n}\exp\left[-(Y-\beta)^2\right]{\rm d}Y=(2{\rm
i})^{-n}\sqrt{\pi}H_n({\rm i}\beta),
\end{equation}
see, for example, \cite{Ryzhik}. In particular
\begin{eqnarray}\label{vzero}
V_{0000}&=&S\int\int {\rm d}\xi_e{\rm
d}\xi_h\psi_{e0}(\xi_e)\psi_{h0}(\xi_h)\exp\left[-v\left(z_e-z_h\right)^2\right]\psi_{e0}(\xi_e)\psi_{h0}(\xi_h)\nonumber\\
&=&S\alpha_e\alpha_h\int\int{\rm d}z_e{\rm
d}z_h\psi^2_{e0}(\alpha_e z_e-a_e)\psi^2_{h0}(\alpha_h
z_h-a_h)\nonumber\\
&=&S\alpha_e\alpha_hN_{e0}^2N_{h0}^2\frac{\alpha_e^2}{\alpha_e^2+\alpha_h^2}\exp(-a_e^2-a_h^2)\int\int{\rm
d}X{\rm d}Y\exp(-c_1X^2-c_2X)\exp(-c_3Y^2-c_4Y)\nonumber
\end{eqnarray}
where
\begin{eqnarray}
c_1=\frac{\alpha_e^2\alpha_h^2}{\alpha_e^2+\alpha_h^2}+v&&
c_2=-2\frac{\alpha_e\alpha_h}{\alpha_e^2+\alpha_h^2}\left(a_e\alpha_h+a_h\alpha_e\right)\\
c_3=\frac{\alpha_e^4}{\alpha_e^2+\alpha_h^2},&&c_4=-2\frac{\alpha_e^2}{\alpha_e^2+\alpha_h^2}\left(a_e\alpha_e-
a_h\alpha_h\right).\end{eqnarray} Finally
\begin{eqnarray}
V_{0000}&=&S\frac{\alpha_e^3\alpha_h}{\alpha_e^2+\alpha_h^2}\frac{\pi}{\sqrt{c_1c_3}}N_{e0}^2N_{h0}^2
\exp\left(\frac{c_2^2}{4c_1}+\frac{c_4^2}{4c_3}-a_e^2-a_h^2\right)\nonumber\\
&=&S\frac{\alpha_e^3\alpha_h}{\alpha_e^2+\alpha_h^2}\frac{1}{\sqrt{c_1c_3}}
\exp\left(\frac{c_2^2}{4c_1}+\frac{c_4^2}{4c_3}-a_e^2-a_h^2\right).\end{eqnarray}
\section{The solution of the constitutive equation}
As we noticed in Ref. \cite{EPJ.2015}, in order to account the
lowest exciton state we take the single function
\begin{equation}\label{excitoneigenfunction}
\psi_0(\rho,\phi)=\frac{\sqrt{2\lambda}}{\sqrt{2\pi}}e^{-\lambda\rho^2/2},
\end{equation}
being the normalized eigenfunction of the Schr\"{o}dinger equation
with the Hamiltonian
\begin{equation}\label{stpar_hamiltonianzero}
H_0=-\frac{{\rm d}^2}{{\rm d}\rho^2}-\frac{1}{\rho}\frac{\rm
d}{{\rm d}\rho}-v_{0000}e^{-\varpi \rho^2}=\hat{\bf
p}_\rho^2-v_{0000}e^{-\varpi \rho^2}
\end{equation}
where $v_{0000}=V_{0000}/R^*, \varpi=a^{*2}w$. Note that now
$V_{0000}$ depends on the field strength $F$ and so also the
corresponding eigenvalue $\varepsilon_0$. We put the function
(\ref{excitoneigenfunction}), with the computed value $\lambda$,
into the expansion (\ref{stpar_rozwiniecie1})
\begin{equation}\label{stpar_rozwiniecie2}
Y\left(\rho,z_e,z_h\right)=\psi_0(\rho)\sum\limits_{j,n=0}^{N}\psi_{ej}(z_e)\psi_{nh}(z_h)Y_{jn},
\end{equation}
where now $Y_{jn}$ are constant coefficients. Equation
(\ref{stpar_konstytuwn3}) takes now the form

\begin{equation}\label{stpar_konstytuwn4}
\left(E_{rs}+\frac{\hat{\bf p}_{\hbox{\boldmath
$\rho$}}^2}{2\mu_{\parl
}}\right)\psi_0(\rho)Y_{rs}-e^{-\varpi\rho^2}\psi_0(\rho)\sum\limits_{nj}V_{rsnj}Y_{nj}=E\frac{M_0}{2\pi\rho_0}\langle
\hbox{e}r\vert\delta(z_e-z_h)\hbox{h}s\rangle\delta\left(\rho-\rho_0\right).
\end{equation}
\noindent After rescaling the spatial variable $\rho\rightarrow
\rho/a^*$  we obtain from (\ref{stpar_konstytuwn4}) the relation
\begin{equation}\label{stpar_konstytuwn5}
\left(k_{rs}^2+\hat{\bf
p}_\rho^2\right)\psi_0(\rho)Y_{rs}-e^{-\varpi\rho^2}\psi_0(\rho)\sum\limits_{nj}v_{rsnj}Y_{nj}=\frac{2\mu_\parl}{\hbar^2}E\frac{M_0}{2\pi\rho_0}\langle
\hbox{e}r\vert\delta(z_e-z_h)\hbox{h}s\rangle\delta\left(\rho-\rho_0\right),
\end{equation}
which, using the quantities
$k_{rs}^2=\frac{E_{rs}}{R^*},v_{rsnj}=\frac{V_{rsnj}}{R^*}$ can be
written as
\begin{eqnarray}\label{stpar_konstytuwn6}
&&\left(k_{rs}^2+\hat{\bf p}_\rho^2-v_{0000}e^{-\varpi\rho^2}\right)\psi_0(\rho)Y_{rs}+v_{0000}e^{-\varpi\rho^2}\psi_0(\rho)Y_{rs}\nonumber\\
&&-e^{-\varpi\rho^2}\psi_0(\rho)\sum_{nj}v_{rsnj}Y_{nj}=\frac{2\mu_\parl}{\hbar^2}E\frac{M_0}{2\pi\rho_0}\langle
\hbox{e}r\vert\delta(z_e-z_h)\hbox{h}s\rangle\delta\left(\rho-\rho_0\right),
\end{eqnarray}
\noindent and, in consequence,
\begin{equation}\label{stpar_konstytuwnuklad1}
\left(k_{rs}^2+\epsilon_0\right)Y_{rs}+v_{0000}\frac{\lambda}{\lambda+\varpi}Y_{rs}
-\frac{\lambda}{\lambda+\varpi}\sum\limits_{nj}v_{rsnj}Y_{nj}=\frac{2\mu_\parl}{\hbar^2}E{M_0}\langle
\hbox{e}r\vert\delta(z_e-z_h)\hbox{h}s\rangle\psi_0\left(\rho_0\right).
\end{equation}
\noindent We obtained a system of linear algebraic equations for
the coefficients $Y_{nj}$. Having them, we determine the amplitude
 $Y$ (or amplitudes, when accounting the heavy- and light hole excitons H and L. Given the amplitude, we compute the polarization inside
 the WPQW. For the further calculations we introduce dimensionless
 quantities $\mathcal{Y}_{rs}$
\begin{equation}
\frac{2M_0}{\epsilon_0\epsilon_b\pi a^*}Y_{rs} =
\mathcal{Y}_{rs}\cdot E
\end{equation}


\noindent and arrived to the formula 
\begin{equation}\label{stpar_niemianowane}
\left(k_{rs}^2 + \epsilon_0\right)\mathcal{Y}_{rs} +
v_{0000}\frac{\lambda}{\lambda + \varpi}\mathcal{Y}_{rs} -
\frac{\lambda}{\lambda + \varpi}\sum_{nj}v_{rsnj}\mathcal{Y}_{nj}
= \frac{\Delta_{LT}}{R^*}\langle
\hbox{e}r\vert\delta(z_e-z_h)\hbox{h}s\rangle\psi_0\left(\rho_0\right),
\end{equation}
where we used the relation
$2\frac{2\mu_\parl}{\hbar^2}\frac{M_0^2}{\epsilon_0\epsilon_b\pi
a*} = \frac{\Delta_{LT}}{R*},$
with $\Delta_{LT}$ being the longitudinal-transversal splitting
energy, (see for example,\cite{RivistaGC}).

\begin{figure}[ht]
\begin{subfigure}[b]{0.5\textwidth}
\includegraphics[width=1.0\linewidth]{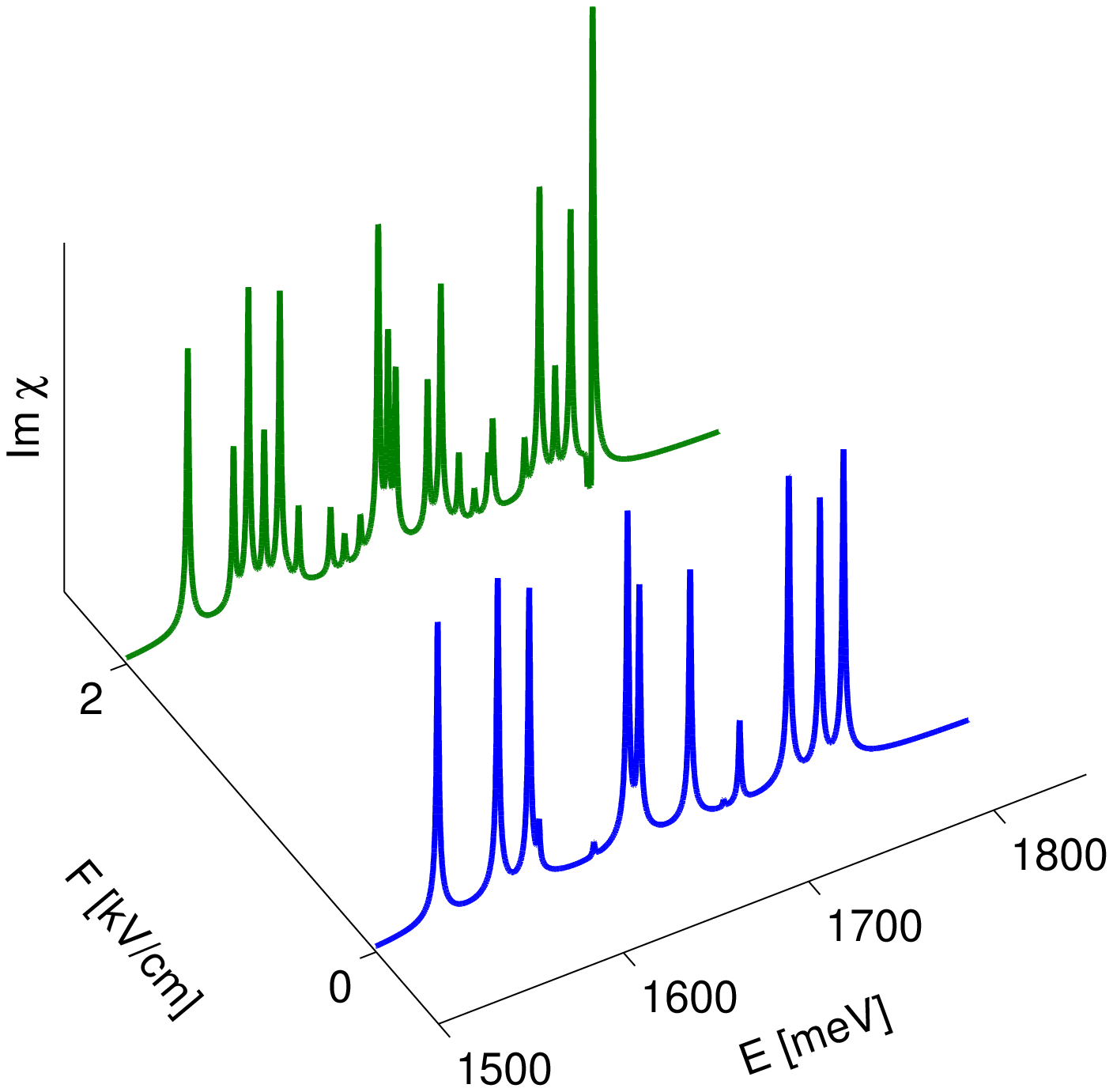} \caption{}
\end{subfigure}
\begin{subfigure}[b]{0.5\textwidth}
\includegraphics[width=1.0\linewidth]{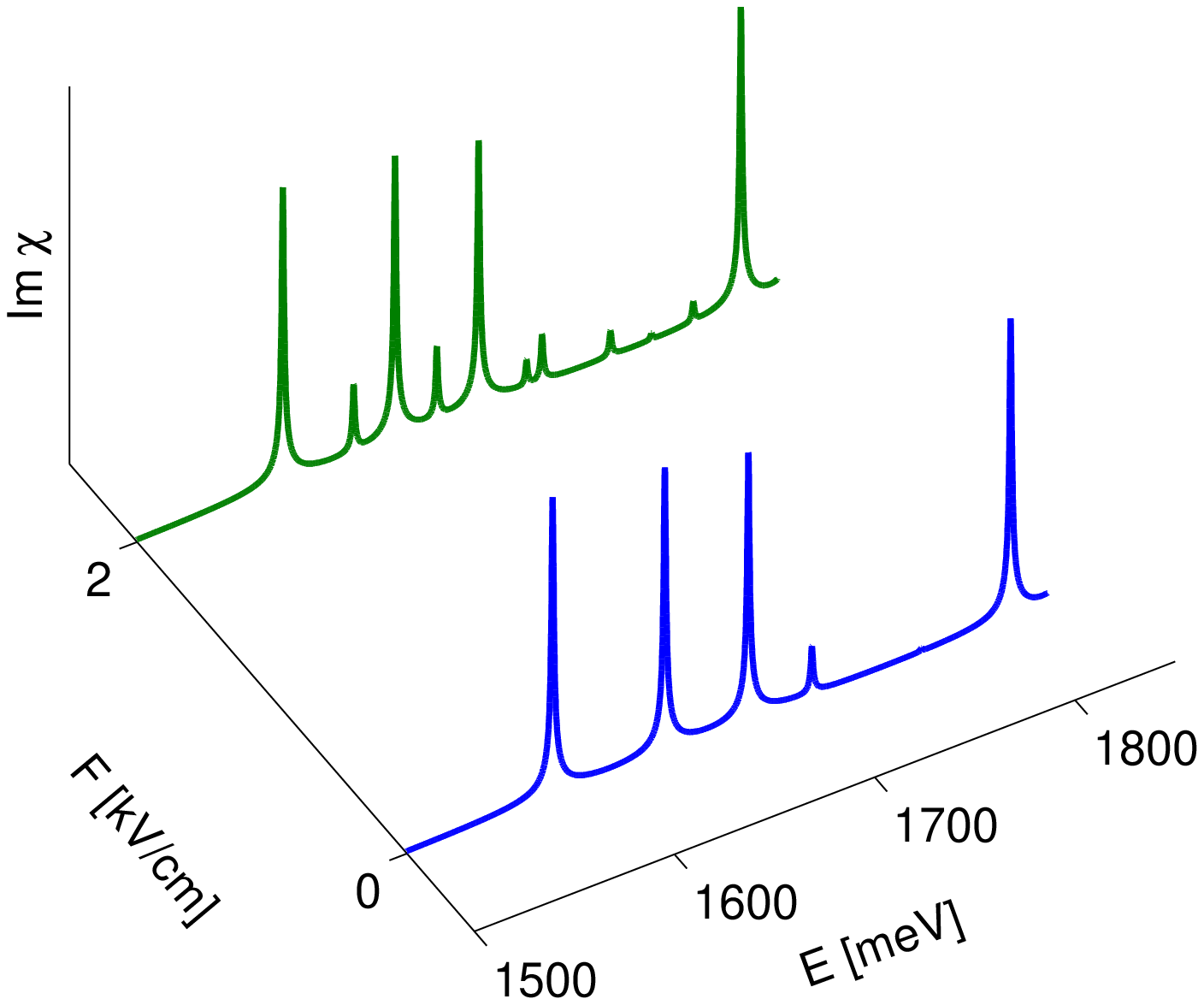} \caption{}
\end{subfigure}
 \caption{\small The imaginary part of the mean susceptibility
form the eq. (\ref{stpar_sr_podatn}), for two
GaAs/Ga$_{0.7}$Al$_{0.3}$As WPQWs, a) of thickness 51 nm, b) of
thickness 32,5 nm } \label{Fig1}
\end{figure}

\section{Results for
G\lowercase{a}A\lowercase{s}/G\lowercase{a}$_{1-
\lowercase{x}}$A\lowercase{l}$_{\lowercase{x}}$A\lowercase{s}
Parabolic Quantum Well and discussion}\label{secIV} 
\begin{figure}[ht!]
\centering
\includegraphics[width=0.5\linewidth]{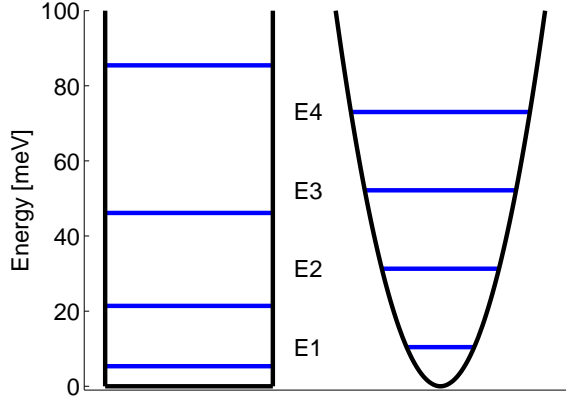}\caption{Comparison of the lowest energy levels in an infinite and parabolic QW, calculated from equation~(\ref{starkshift}), for GaAs(Well)/Ga$_{0.7}$Al$_{0.3}$As (Barrier) and thickness $L=32.5 \hbox{nm}$.}\label{studnie}
\end{figure}

The
calculation of the WPQW electrooptical functions consists of
several steps. First, we define the confinement energies
$\hbar\omega_{e,h}$. To this end we choose a specific WPQW having
in mind the experimental results  of Miller et
al.\cite{MillerGossard}. They obtained optical spectra for
GaAs(Well)/Ga$_{0.7}$Al$_{0.3}$As (Barrier) QWs of three
thicknesses: $L=51 \pm~3,5~\hbox{nm}, L=32,5 \pm~3.5~\hbox{nm},
L=33.6 \pm~3.5~\hbox{nm}$. We have performed the calculations for
the thicknesses $L=51~\hbox{nm}$ and  $L=32,5 ~\hbox{nm}$. The
confinement parameters  were obtained from the lowest energy
levels of equivalent rectangular QWs with confinement potentials
$V_{e,conf}=410.38~\hbox{meV},
V_{h,conf}=72.42~\hbox{meV}$~\cite{arxiv}.\\ 
The QWs energy states were obtained by standard methods (see, for example, Ref.
\cite{Davies} and \cite{arxiv} for calculation details), using the
band parameters from Table \ref{parametervalues}. In the considered parabolic QW, the lowest energy levels are very similar to the case of an infinite, rectangular well, as one can see on the Fig.~\ref{studnie}

The values
$a^*$, $R^*$ are appropriate for electrons and holes for the QW
material, and are defined as
\begin{equation}\label{rydberg}
 R^*=\frac{me^4}{2(4\pi\epsilon_0\epsilon_b)^2\hbar^2},\,\,\,\,\,\,\,\,
 a^*=\frac{\hbar^2(4\pi\epsilon_0\epsilon_b)}{m e^2}.
 \end{equation}
 The corresponding values, listed in Table \ref{parametervalues}, were obtained by using in
 (\ref{rydberg}) the appropriate effective masses: $m_e$ for
 $R^*_e,a^*_e$, and $\mu_{\parallel H,L}$ for $R^*_H, a^*_H$ and
 $R^*_L, a^*_L$; $\mu_{\parallel H,L}$ are the in-plane reduced masses
 for the electron-hole pair and for the heavy- and light-hole
 exciton data.

\begin{figure}[ht]
\begin{subfigure}[b]{0.5\textwidth}
\includegraphics[width=1.0\linewidth]{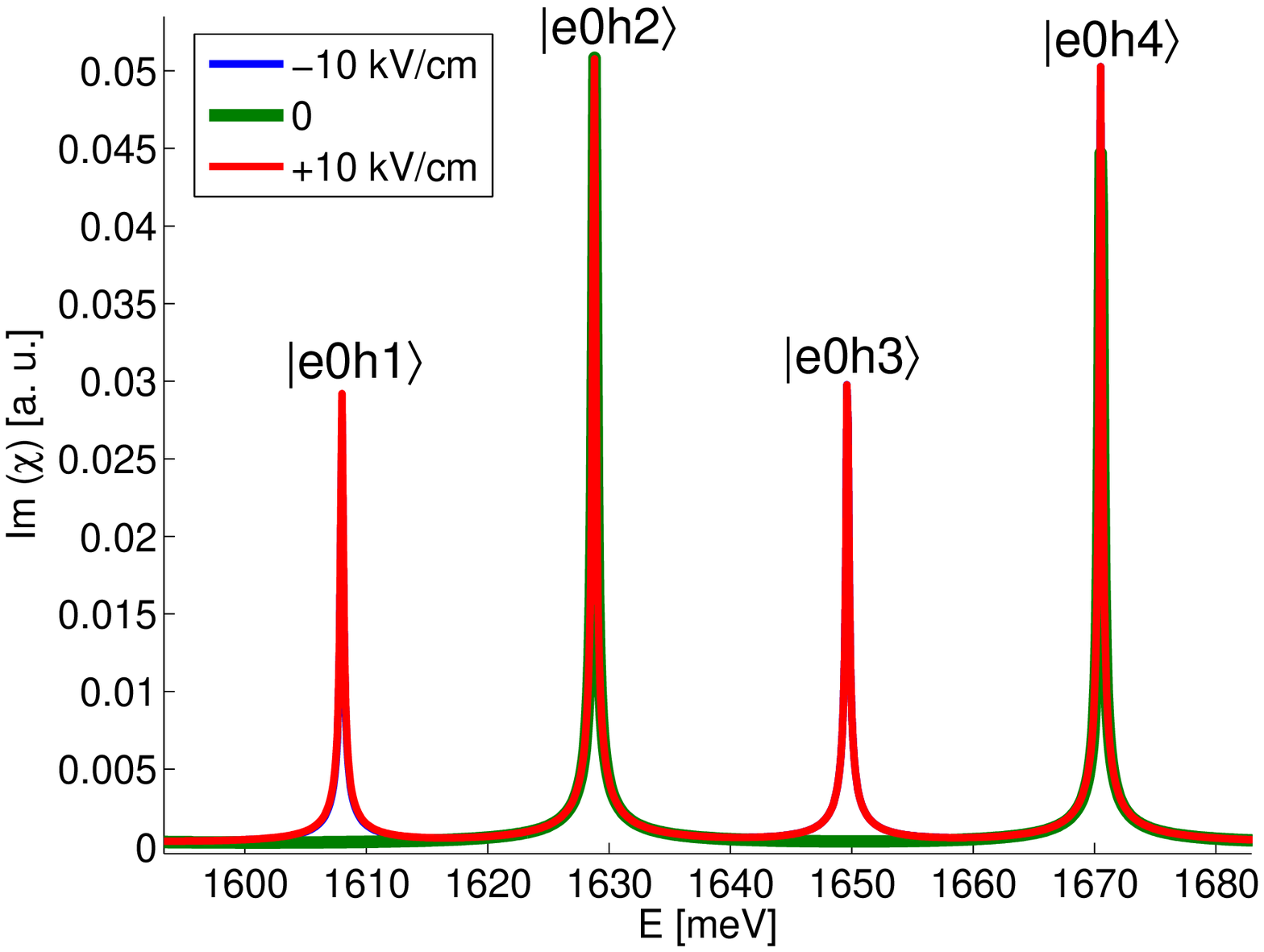}\caption{}
\end{subfigure}
\begin{subfigure}[b]{0.5\textwidth}
\includegraphics[width=1.0\linewidth]{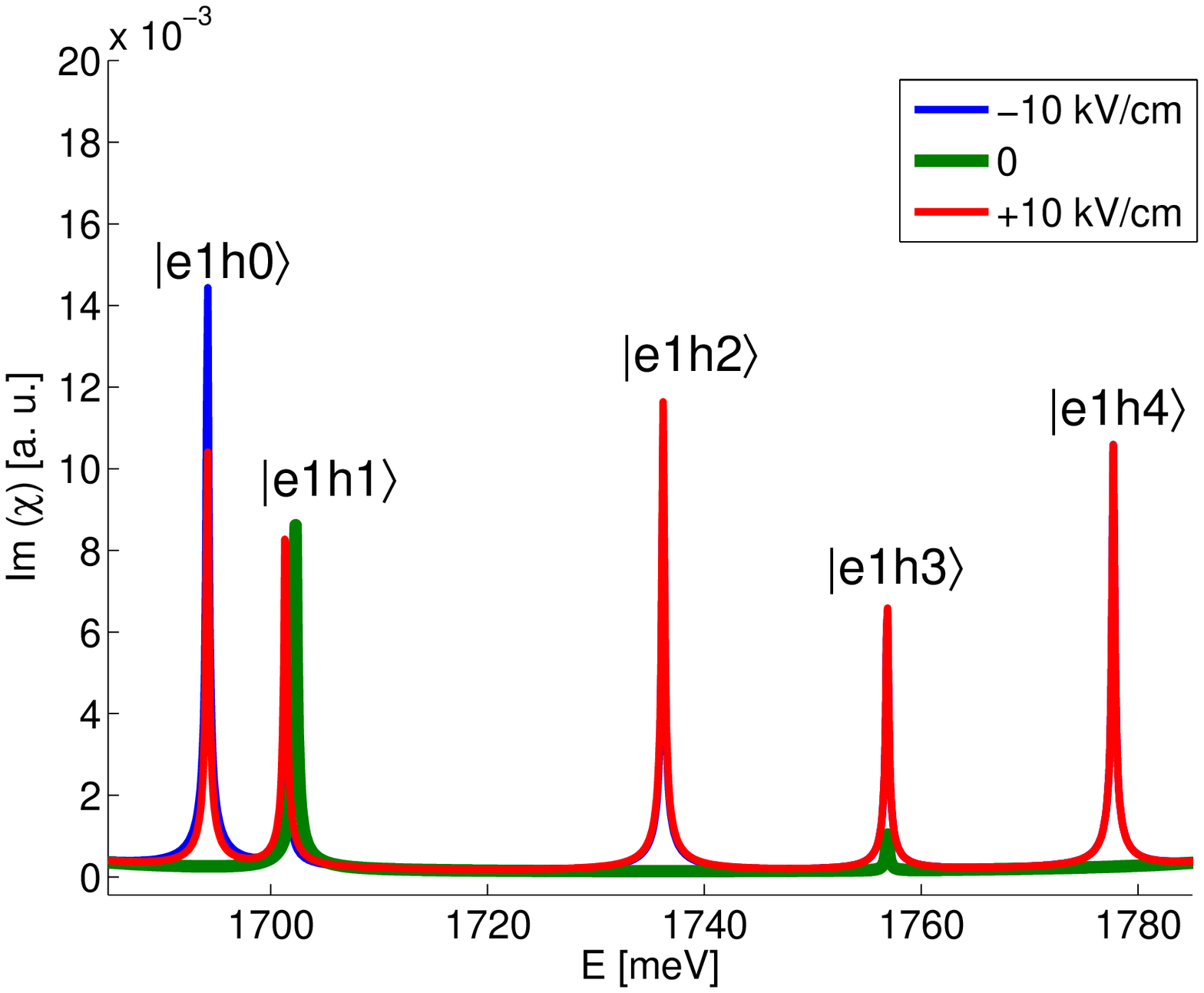}\caption{}
\end{subfigure}
\caption{\small The imaginary part of the mean
electrosusceptibility for the GaAs/Ga$_{0.7}$Al$_{0.3}$As WPQW of
thickness 32.5 nm, for different energy intervals}
\label{Fig2}
\end{figure}
\begin{figure}[ht]
\begin{subfigure}[b]{0.5\textwidth}
\includegraphics[width=1.0\linewidth]{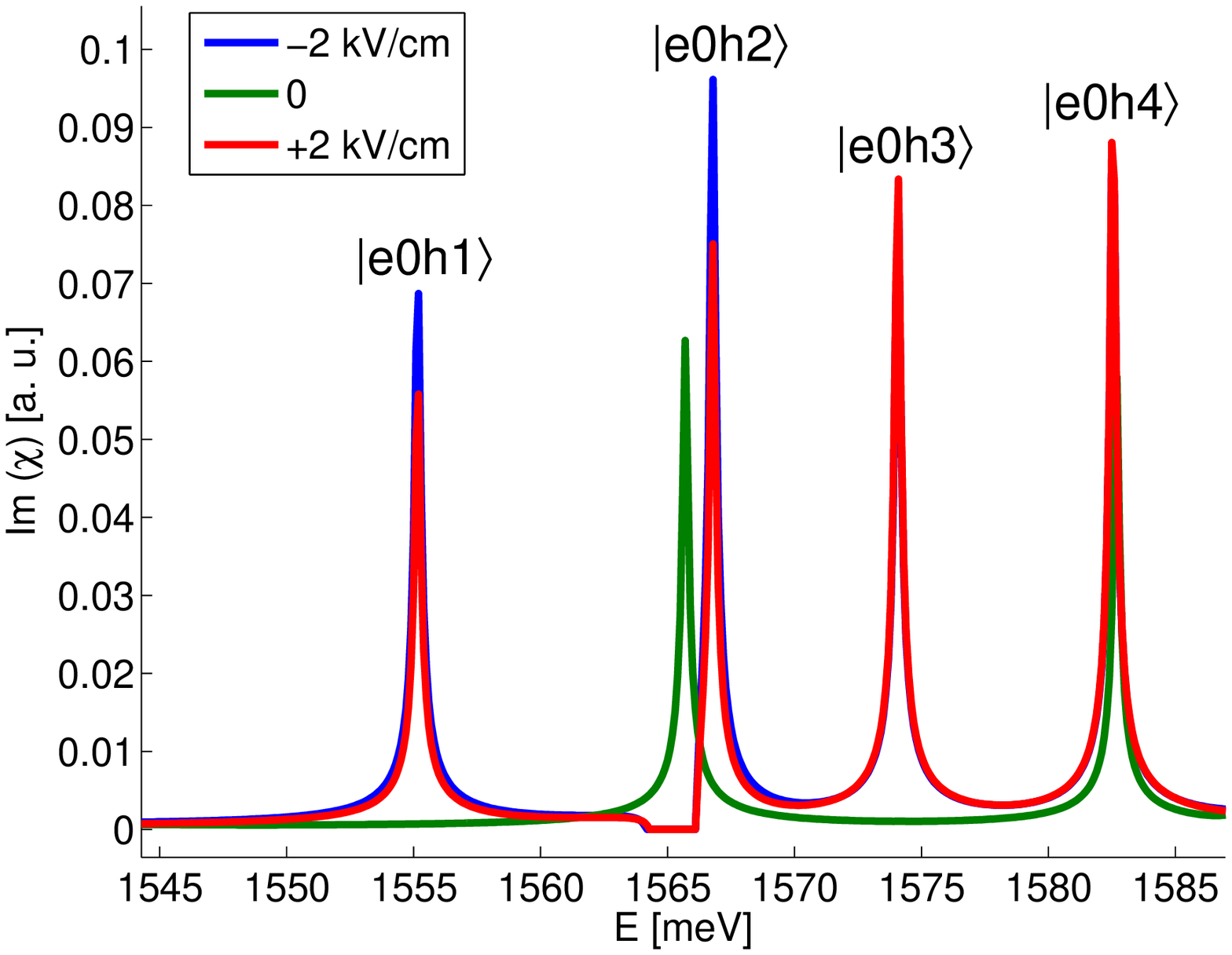}\caption{}
\end{subfigure}
\begin{subfigure}[b]{0.5\textwidth}
\includegraphics[width=1.0\linewidth]{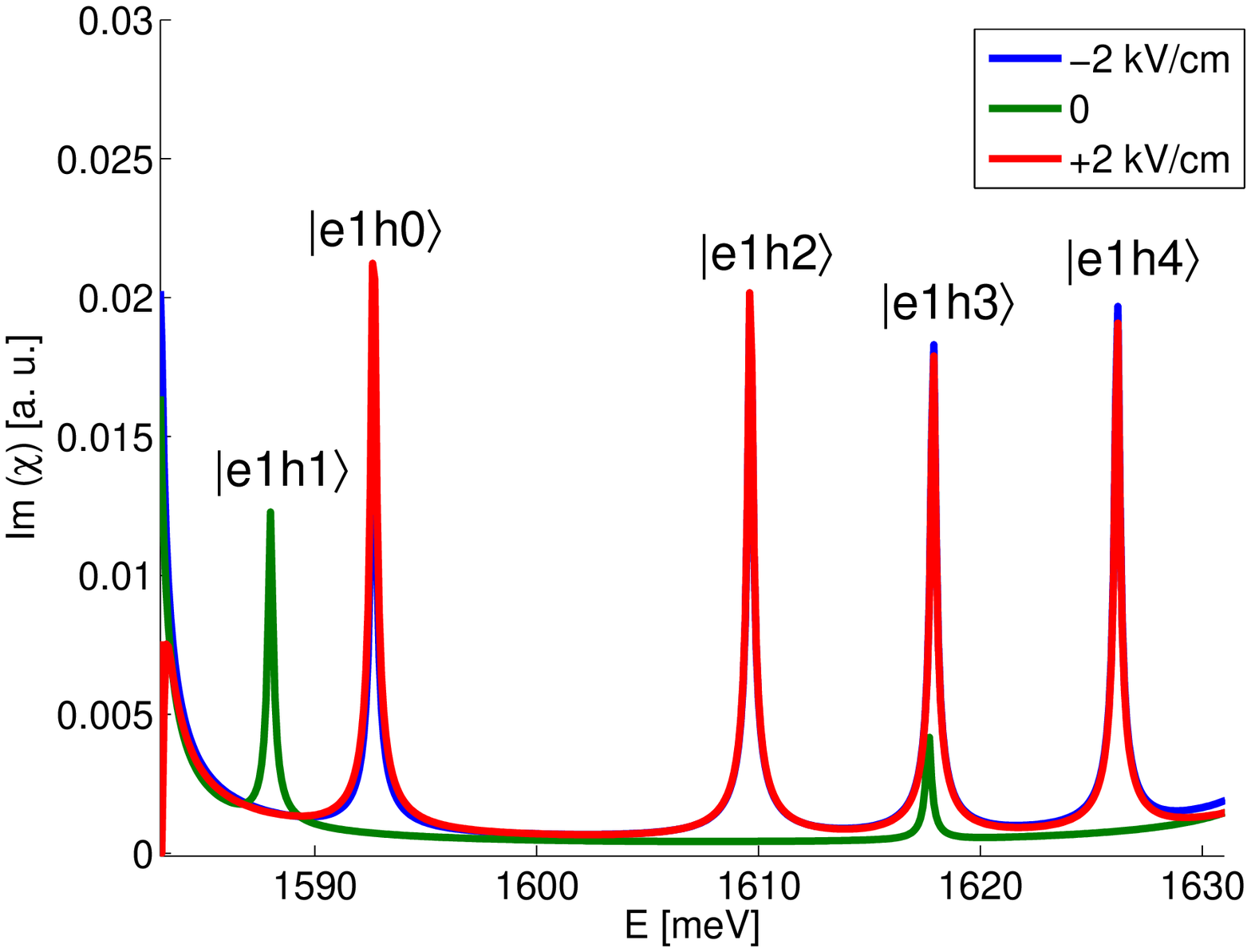}\caption{}
\end{subfigure}
\caption{\small The same as in Fig.~\ref{Fig2}, for the WPQW of
the thickness  51 nm} \label{Fig3}
\end{figure}

The results for the confinement energy states are displayed in
Table \ref{confineparamers}. From this energies we obtained the
confinement energies as
\begin{equation}
\hbar \omega_e=2E_{e0}, \qquad \hbar \omega_{hH,L} = 2E_{0zH,L}.
\end{equation}

\begin{table}[ht]
\caption{Confinement parameters for the WPQWs from Ref.
\cite{MillerGossard}, dimensions in nm, energies in meV,}
\label{confineparamers}
\begin{center}
\begin{tabular}{c c c cc cccccc}
\hline\\
$L$ &$\hbar\omega_e$  & $\hbar\omega_{hH}$ & $\hbar\omega_{hL}$&$E_{e0}$&$E_{0zH}$&$E_{0zL}$&$\alpha_ea_H^*$&$\alpha_{hH}a_H^*$&$\alpha_{hL}a_L^*$\\
\hline
32.5 & 81.66&19.74  &108.4  &40.83&9.87&54.2&4.21&4.95&4.76\\
51.5&43.56&8.46&34.4&21.78&4.23&17.2&3.07&3.08&2.68\\
\hline\\
\end{tabular}
\end{center}
\end{table}

\begin{table}[ht]
\begin{small}
\caption{ \small Confinement states accounted in
computation}\label{Confinement.states}
\begin{center}

\begin{tabular}{|c|c|c|c|c|} \hline
 $\vert e0h0\rangle \rightarrow \vert 1\rangle$& $\vert e0h1\rangle  \rightarrow \rangle\vert 2\rangle$&$\vert e0h2\rangle
 \rightarrow \vert 3\rangle$&$\vert e0h3\rangle \rightarrow \vert 4\rangle$&$\vert e0h4\rangle \rightarrow \vert 5\rangle$\\
\hline $\vert e1h0\rangle \rightarrow \vert 6\rangle$ &$\vert
e1h1\rangle \rightarrow \vert 7\rangle$
&$\vert e1h2\rangle \rightarrow \vert 8\rangle$&$\vert e1h3\rangle \rightarrow \vert 9\rangle$&$\vert e1h4\rangle \rightarrow \vert 10\rangle$\\
\hline $\vert e2h0\rangle \rightarrow \vert 11\rangle$& $\vert
e2h1\rangle \rightarrow \vert 12\rangle$&$\vert e2h2\rangle
\rightarrow \vert 13\rangle$&$\vert e2h3\rangle \rightarrow \vert
14\rangle$&$\vert e2h4\rangle \rightarrow
\vert 15\rangle$\\
\hline $\vert e3h0\rangle \rightarrow \vert 16\rangle$& $\vert
e3h1\rangle \rightarrow \vert 17\rangle$&$\vert e3h2\rangle
\rightarrow \vert 18\rangle$&$\vert e3h3\rangle \rightarrow
\vert 19\rangle$&$\vert e3h4\rangle \rightarrow \vert 20\rangle$\\
\hline $\vert e4h0\rangle \rightarrow \vert 21\rangle$&$\vert
e4h1\rangle \rightarrow \vert 22\rangle$&$\vert e4h2\rangle
\rightarrow
\vert 23\rangle$&$\vert e4h3\rangle \rightarrow \vert 24\rangle$&$\vert e4h4\rangle \rightarrow \vert 25\rangle$\\
\hline
\end{tabular}
\end{center}
\end{small}
\end{table}
Using the above parameters and taking into account the lowest 25
confinement states (see Table \ref{Confinement.states}) we have
solved the eqn. (\ref{stpar_niemianowane}) and obtained the
coefficients $Y_{jn}$ from which we have determined the induced
polarization inside the WPQW by the relation

\begin{equation}
P(z) = 2M_0\psi_0(\rho_0)\sum\limits_{j,n=0}^N\vert
\hbox{e}j\hbox{h}n\rangle(z)Y_{jn},
\end{equation}
with the notation
\begin{equation}
\vert \hbox{e}j\hbox{h}n\rangle(z) = \psi_{ej}(z)\psi_{hn}(z).
\end{equation}
Having the polarization, we compute the mean dielectric
susceptibility
\begin{equation}\label{stpar_sr_podatn}
\overline{\chi} =
\pi\epsilon_b\psi_0(\rho_0)\sum\limits_{\ell=0}^N
\mathcal{Y}_\ell\langle 1\vert \ell\rangle_{\Lambda/2}
\end{equation}
where $\langle 1\vert\ell\rangle_{\Lambda/2} =
\frac{1}{\Lambda}\int\limits_{-\Lambda/2}^{\Lambda/2}\vert\ell\rangle(\zeta){\rm
d}\zeta,  \Lambda=\frac{L}{a^*}$, and the index $\ell$ runs over
the 25 states listed in Table~\ref{Confinement.states}.
Then, having the mean susceptibility, one can compute, using the
appropriate boundary conditions, the optical functions
(reflectivity, transmission, and absorption).
\begin{figure}[ht]
\begin{subfigure}[b]{0.5\textwidth}
\includegraphics[width=1\linewidth]{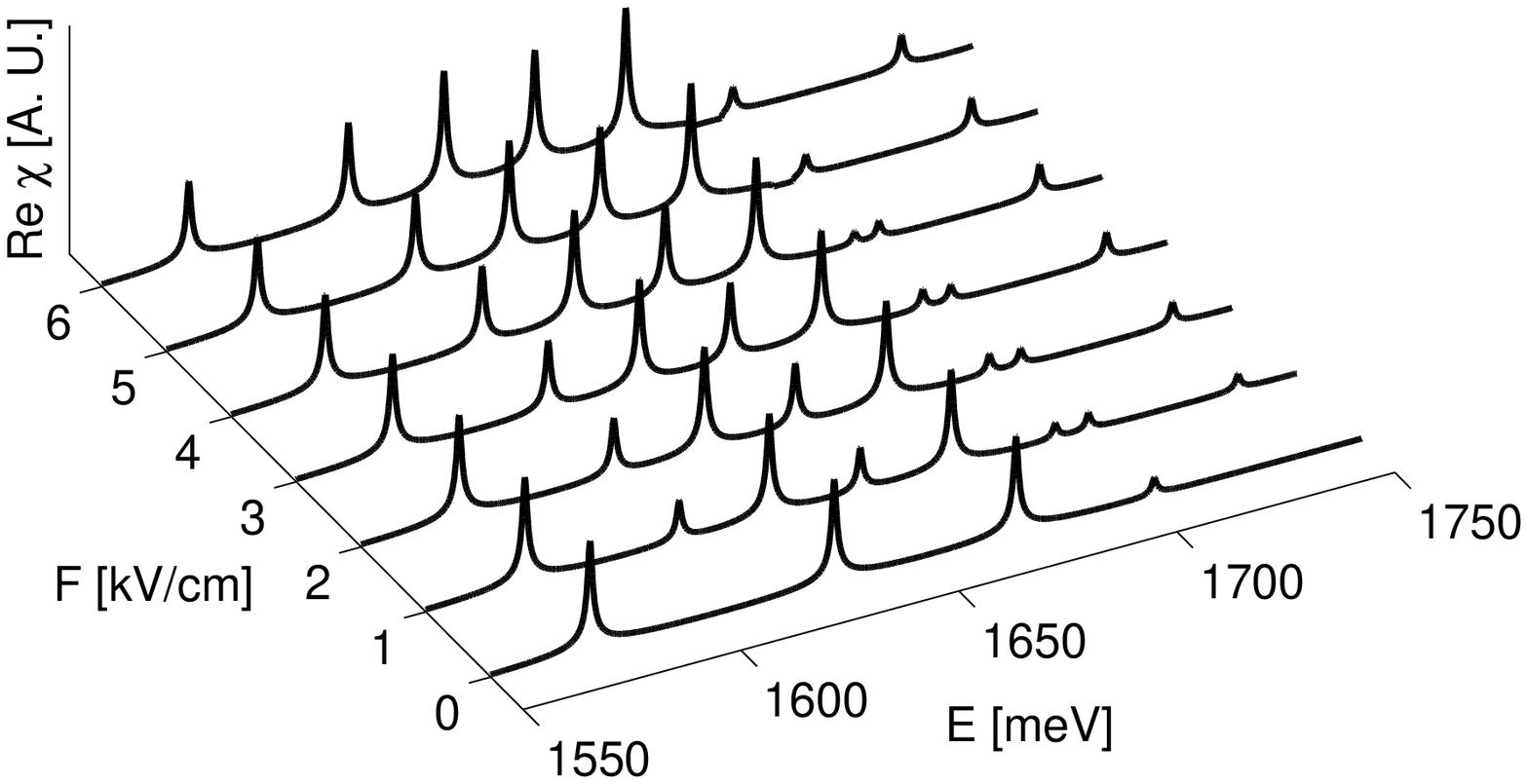}
\end{subfigure}
\begin{subfigure}[b]{0.5\textwidth}
\includegraphics[width=1\linewidth]{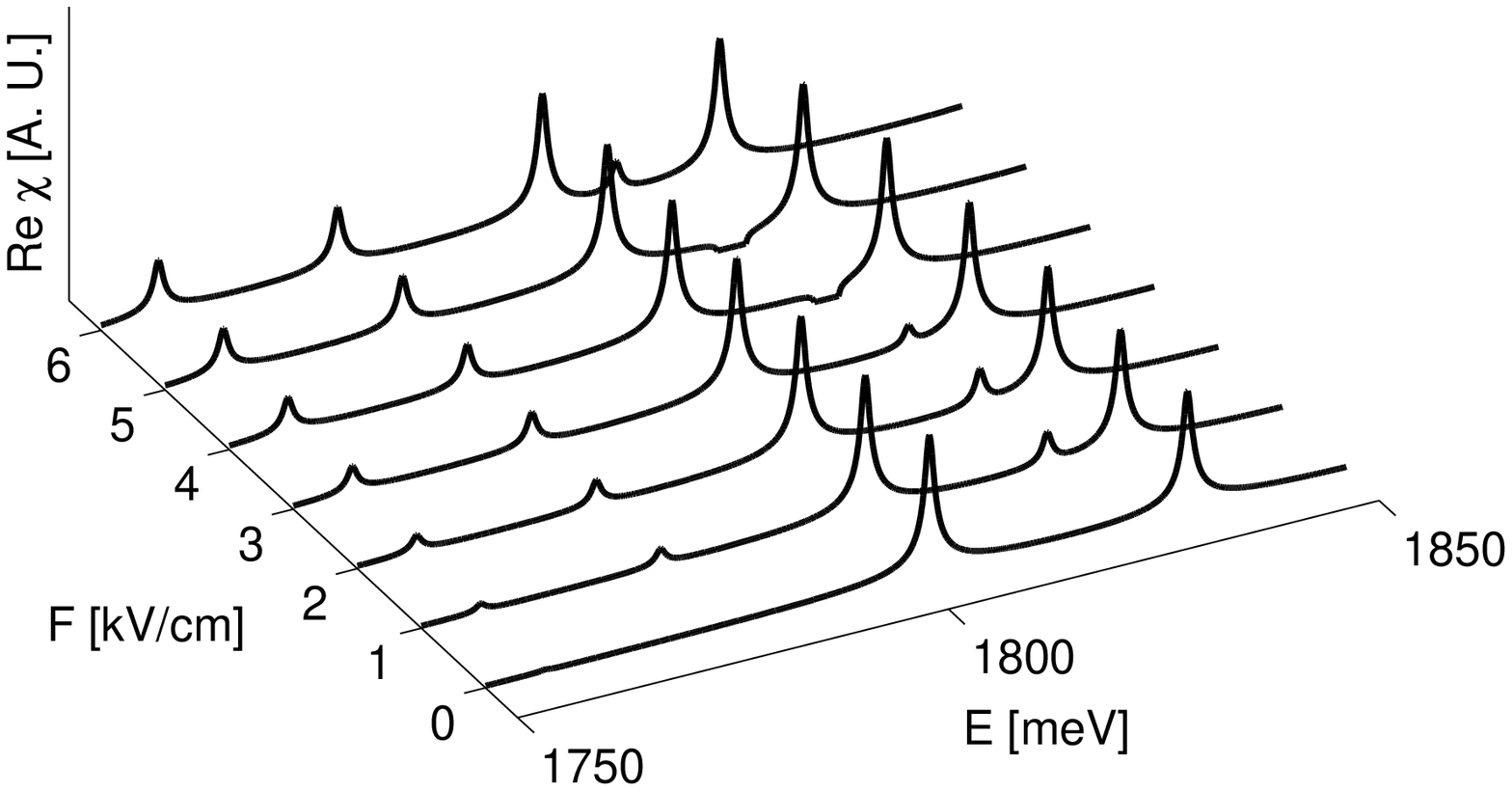}
\end{subfigure}
 \label{Fig4}
 \caption{\small The real part of the mean electrosusceptibility
for the GaAs/Ga$_{0.7}$Al$_{0.3}$As WPQWs of thicknesses 32.5 nm
 for different energy intervals and applied
field strengths}
\end{figure}
 We have computed the electrosusceptibility for two WPQWs of
thicknesses $32.5~\hbox{nm}$ and $51~\hbox{nm}$. The parameters
$S=2.6,\varpi=0.154$ and $v=0.5$ were determined with the procedure
described in Ref.~\cite{EPJ.2015}.
\begin{figure}[ht]
\begin{subfigure}[b]{0.5\textwidth}
\includegraphics[width=1\linewidth]{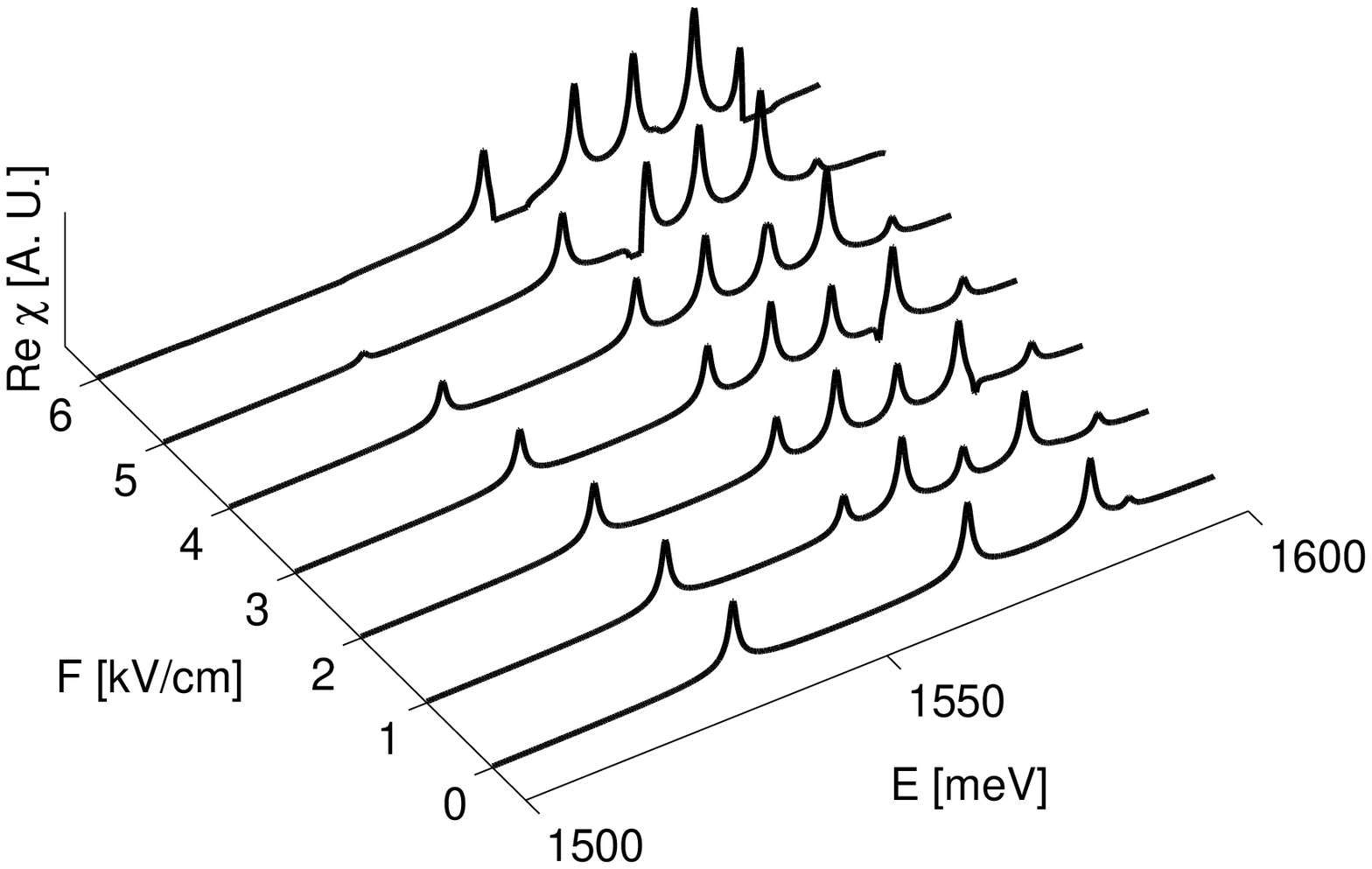}\caption{}
\end{subfigure}
\begin{subfigure}[b]{0.5\textwidth}
\includegraphics[width=1\linewidth]{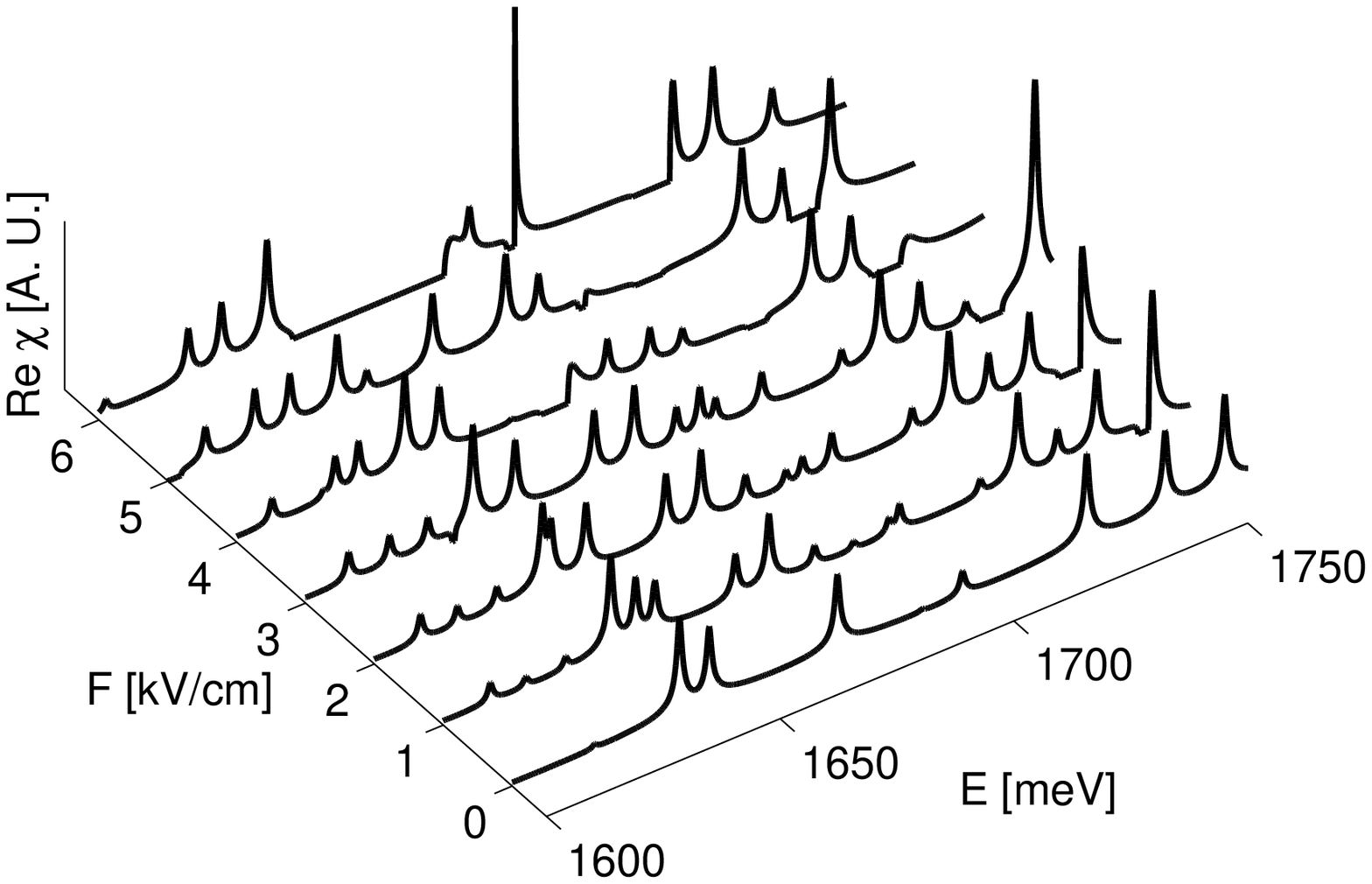}\caption{}
\end{subfigure}
\begin{subfigure}[b]{0.5\textwidth}
\includegraphics[width=1\linewidth]{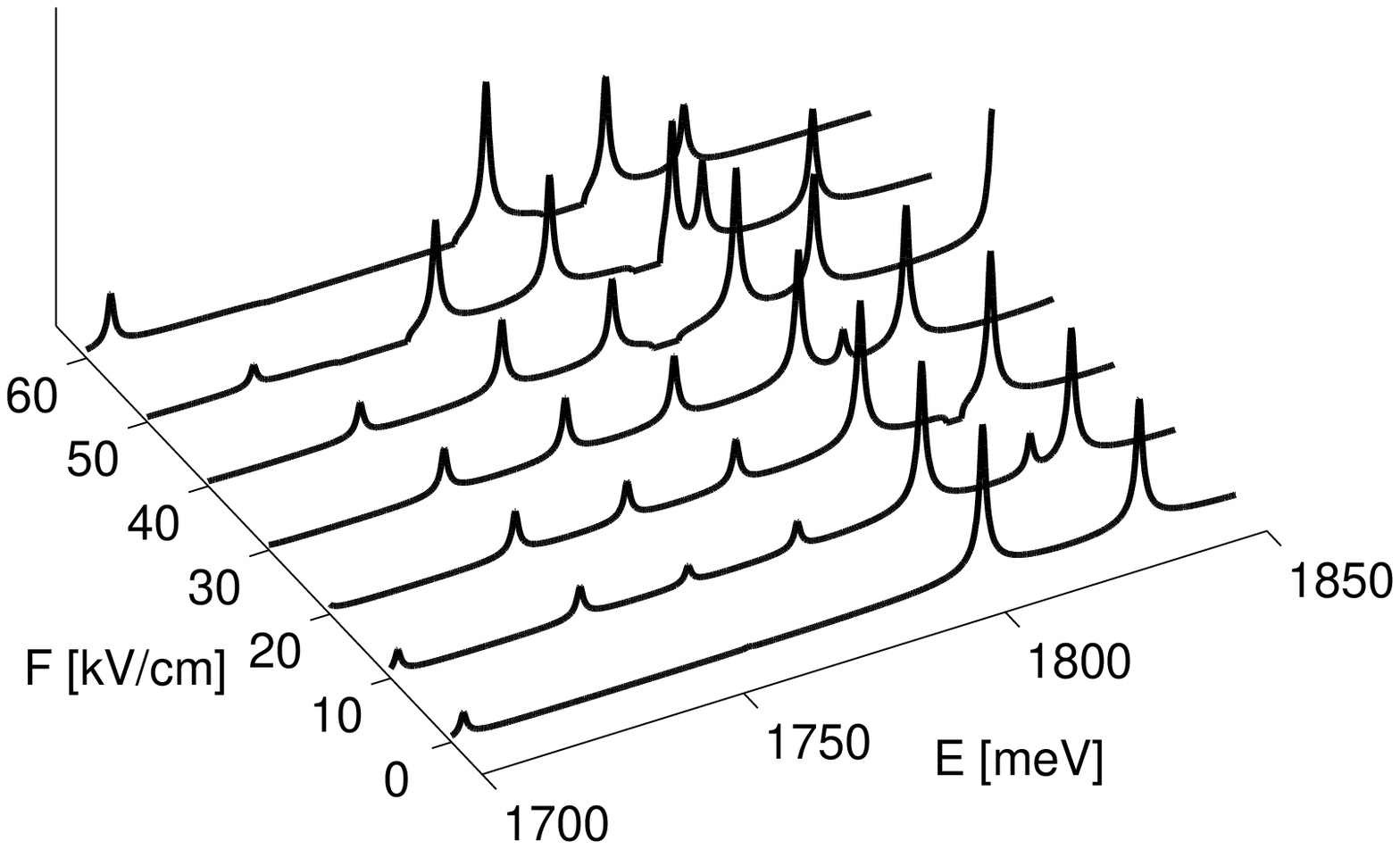}\caption{}
\end{subfigure}
\begin{subfigure}[b]{0.5\textwidth}
\centering
\includegraphics[width=0.9\linewidth]{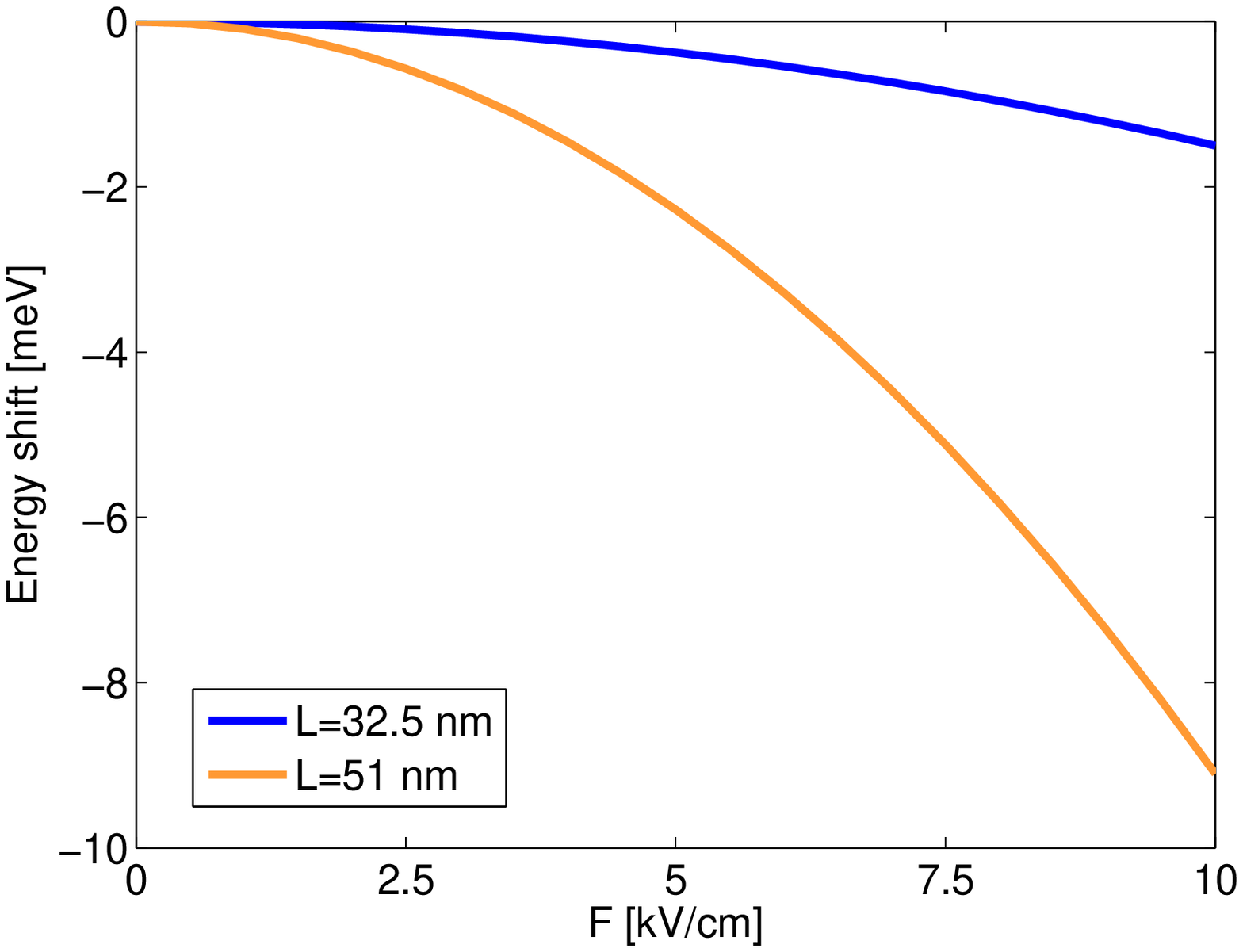}\caption{}\label{Fig7}
\end{subfigure}

\caption{\small a), b) The real part of the mean electrosusceptibility
for the GaAs/Ga$_{0.7}$Al$_{0.3}$As WPQWs of thickness
  51 nm , for different energy intervals and applied
field strengths. c) the impact of high electric fields (WPQW of
thickness 32.5 nm) d) The comparison of the Stark energy shift of the
two considered WPQWs for the lowest resonance.} \label{Fig5}
\end{figure}
Assuming a certain value of the coherence radius $\rho_0$, we have
determined the lowest excitonic eigenfunction $\psi_0$. Finally,
taking a certain value of the damping parameter $\mit\Gamma$, we
have solved the constitutive equation (\ref{stpar_konstytuwn3}),
obtaining the coherent amplitudes. From the amplitudes we have
computed the mean dielectric susceptibility
(\ref{stpar_sr_podatn}). The advantage of the RDMA is that we
obtain simultaneously the real and the imaginary part of the
susceptibility. The results for the the imaginary part of the mean
susceptibility of the considered WPQWs are displayed in Fig.1,
Fig.2, and Fig.3. In Fig. \ref{Fig1}  we show the general effect
of the applied electric field for two GaAs/Ga$_{0.7}$Al$_{0.3}$As
WPQws. We observe the red shift of the resonances, changes in the
oscillator strengths, and the occurrence of new resonances due to
the broken symmetry. The spectra for $F=0$ agree well with the
experimental results by Miller \emph{et. al. }
\cite{MillerGossard} and our previous theoretical results
\cite{EPJ.2015}. In Fig. 2 and Fig. 3 we show the obtained spectra
in a more detailed form, as compared to Fig. 1.

\begin{figure}[ht!]
\begin{subfigure}[b]{0.5\textwidth}
\includegraphics[width=1\linewidth]{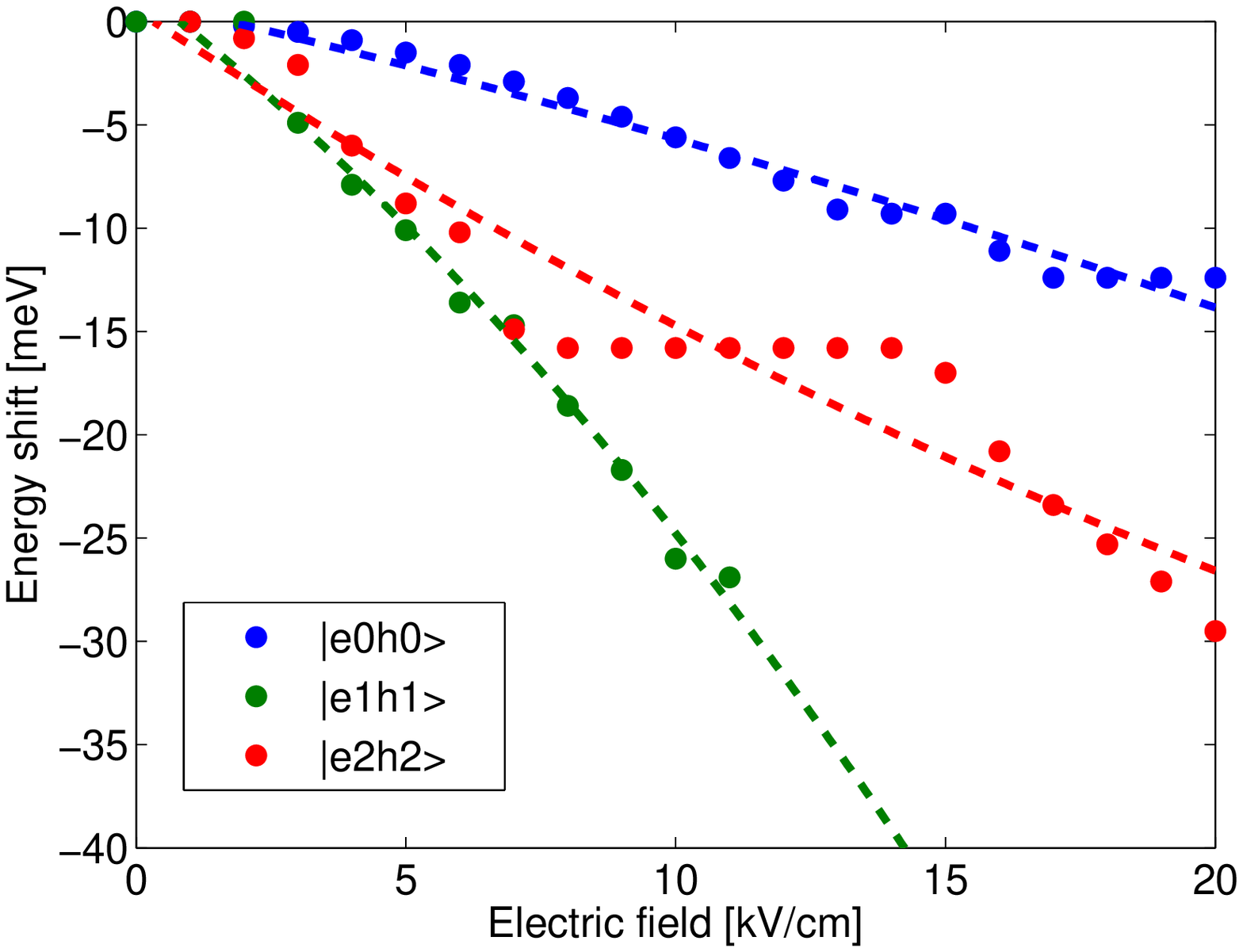}\caption{}
\end{subfigure}
\begin{subfigure}[b]{0.5\textwidth}
\includegraphics[width=1\linewidth]{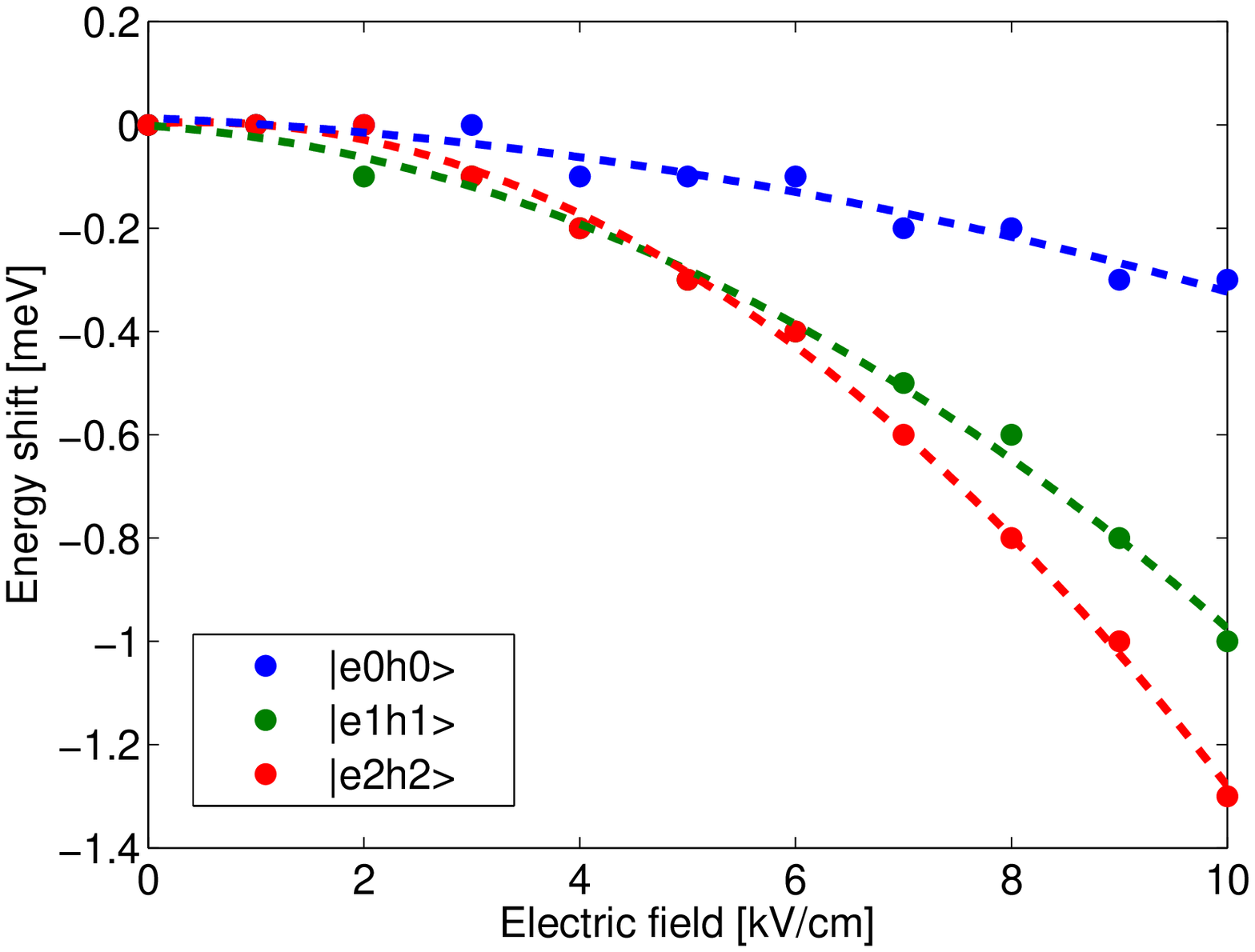}\caption{}
\end{subfigure}
\hfill
\caption{\small The Stark energy shift for the lowest resonances:
(a)for the GaAs/Ga$_{0.7}$Al$_{0.3}$As WPQW of thickness 51 nm and
(b) for the GaAs/Ga$_{0.7}$Al$_{0.3}$As WPQW of thickness 32.5 nm
} \label{Fig6}
\end{figure}

 In all the cases we
observe changes in the placement of resonances, and the occurrence
of new peaks attributed to different symmetries for the electron
and the hole confinement functions. For high values of the applied
electric field the effects are smaller which is due to the
decreasing overlap of the electron and the hole confinement
functions. We also observe that the shape of the spectra changes
with the change of the direction of the applied field. Using the
properties of the RDMA, we also obtained the real part of the mean
electrosusceptibility, which is displayed in Fig.4 and Fig. 5a.
The impact of high electric fields is displayed in Fig.5b, where
we show the changes in the real part of the electrosusceptibility
for the applied fields up to 60 kV/cm. Our method allows to
determine the energy shift as a function of the applied field. We
have computed the energy shift for the lowest confinement states.
We observe the quadratic Stark shift for the lowest state and a
more complicated field-dependence for higher states, as is
displayed in Fig.6. Finally, we show  that the energy shift
drastically depends on the thickness of the QW (Fig.7), as was
also observed for narrow QWs (see, for example, \cite{Polland},
\cite{RivistaGC} and references therein). Since the confinement energy depends on the QW thickness as $L^{-2}$, we see that for rectagular QW, $\Delta E = -CL^4$. The energy shift in the parabolic well on the Fig. \ref{Fig7} also closely follows this relation.

\section{Conclusions}\label{secV}
We have developed a simple mathematical procedure to calculate the
electrooptical functions of Wide Parabolic Quantum Wells. Using
the Real Density Matrix Approach and a model e-h interaction
potential, we derived an analytical formula for the WPQW
electrosusceptibility, from which another electrooptical functions
can be obtained. The presented method has been used to investigate
the electrooptical functions of GaAs/Ga$_{1-x}$Al$_x$As WPQWs for
the case of radiation incidence parallel to the growth direction.
We have obtained the red shift of the resonances, changes in the
oscillator strengths and new peaks related to electronic
transitions forbidden for the case with absent electric field. We
also observed the dependence of the spectra on the size of the QW
and on the direction of the applied field. For the cases where the
experimental data were available (for example, for WPQWs with
\textbf{F}=0), we obtained a good agreement of our theoretical
results with experiment. We hope that our results may stimulate
experiments on Quantum Confined Stark Effect in WPQWs.\end{document}